# Nurses as agents for achieving Environmentally Sustainable Health Systems: A bibliometric analysis

**Olga María Luque-Alcaraz, Msc [1,2,3], Pilar Aparicio-Martinez PhD [3,4*] , Antonio Gomera PhD [2] and Manuel Vaquero PhD [2,3,4]**

[1] Neurosurgery Department, University Hospital Reina Sofia's, Andalusian Health Care System, Cordoba, Spain; olgaluque33@gmail.com (O.M L.)

[2] Environmental Protection Office (SEPA), Campus Rabanales, University of Córdoba, 14014 Córdoba, Spain; olgaluque33@gmail.com (O.M L.), agomera@uco.es (A.G.), en1vaabm@uco.es (M. V-A)

[3] IMIBIC GC 12 Research Groups of Clinical-Epidemiological Research in Primary Care, University Biomedical Program for Occupational Medicine, Occupational Epidemiology and Sustainability, 14071, Cordoba, Spain; n32apmap@uco.es (P.A-M)

[4] Nursing, Pharmacology and Physiotherapy Department, Faculty of Medicine and Nursing, University of Cordoba, Spain; olgaluque33@gmail.com (O.M. L.), n32apmap@uco.es (P.A-M.), en1vaabm@uco.es (M. V-A)

*Correspondence: n32apmap@uco.es (P.A-M.)

**Funding:** This research received no external funding; nevertheless, the project has received an award from the excellent official nursing school, Cordoba, Spain, in 2020.

**Data Availability Statement**: The data presented in this study are available on request from the corresponding author. The data are not publicly available due to privacy restrictions.

**Conflict of interests:** The authors declare that they have no known competing financial interests or personal relationships that could have appeared to influence the work reported in this paper.

**Abstract:**

Objective: To analyze the current scientific knowledge and research lines focused on environmentally sustainable health systems, including the role of nurses during this last decade. Methods: A bibliometric analysis was carried out. The gathering was obtained via three databases (WOS, Scopus, and Pubmed), and the PRISMA recommendations were followed to select bibliometric data. Results: The search resulted in 159 publications, significantly increasing the trends from 2017 to 2021 ($p$=0.028). The most relevant countries in this area were the United States of America, the United Kingdom, and Sweden. Also, the top articles were from relevant journals, indexed in JCR, and the first and the second quartile linked to the nursing field and citations ($p$<0.001). Discussion: The study revealed that the publication trend is growing, but there is a lack of experimental data on creating such systems. Additionally, education is key to achieving environmentally sustainable health systems via institutions and policies.

## Introduction

The environment is a determining factor in the well-being and health of the population, causing a negative impact when it is toxic or unbalanced (Watts et al., 2021). Air and soil pollution, imminent climate change, the destruction of healthy ecosystems, and the creation of ideal ecosystems for new microorganisms, such as the SARS-COVID-2 virus, among other factors, are reducing the quality of life and increasing mortality (Fields et al., 2021). From Florence Nightingale to the present, the patient's environment makes it possible to improve the patient's disease process (Kiang & Behne, 2021). Still, if this is a damaged or contaminated environmental system, it can even harm it in the short and long term health (Fields et al., 2021).

In the health sector, health systems are one of the more significant industries that consume a tremendous amount of water, food, plastic materials, and energy (Fields et al., 2021). World Health Organization (WHO) indicated in 2009 that the health sector might have one of the highest footprints linked to energy and materials consumption (World Health Organization, 2009). Sustainable awareness and climate-friendly programs can provide a high quality of care and reduce the production of waste, plastic, and emissions (Kiang & Behne, 2021). The WHO has promoted such programs and interventions since the beginning of 2010 (General Assembly, 2011), but it was not until 2017 that a definition of environmentally sustainable health systems was given. Environmentally sustainable health systems are "health system that improves, maintains or restores health, while minimizing negative impacts on the environment and leveraging opportunities to restore and improve it, to the benefit of the health and well-being of current and future generations" (World Health Organization, 2017). To achieve such a definition, the pillar is the integration of healthcare workers, especially nurses, since they are the primary healthcare workforce (Álvarez-Nieto et al., 2022). Diverse authors have highlighted the vital role of nursing in climate change and the relevance of nursing education in the environmental awareness (Anåker et al., 2015).

The education in sustainability and its awareness among nurses have been described as pillars to mitigate the negative impact of pollution on people's health (Anåker & Elf, 2014; Kearns & Kearns, 2021). Therefore, nurses are engines of change in the current health system regarding environmental sustainability through research and projects are integrated to achieve it (Lilienfeld et al., 2018; Richardson, Heidenreich, et al., 2016), which has been modified or altered by the pandemic (Fields et al., 2021). Recent research indicated a scarcity of studies and research focused on nurses and environmentally sustainable health systems (Osingada & Porta, 2020). This pandemic has increased hospital waste and disposal and unsustainable options, limiting the sustainable policies instituted (Sarkodie & Owusu, 2021). There seem to be differences between creating interventions focused on environmentally sustainable health systems, including nurses (Álvarez-Nieto et al., 2022; Osingada & Porta, 2020) and the scarcity of research on this topic (Sarkodie & Owusu, 2021). Based on these discrepancies, it is necessary and appropriate to investigate how this revolution toward environmental sustainability is currently going and what is the role of nurses as agents to create environmentally sustainable health systems. The objective of this research was to analyze the current scientific knowledge, and research lines focused on environmentally sustainable health systems, including the role of nurses during this last decade. Also, a secondary objective was to determine the nursing education and interventions to improve the environmental awareness among them.

## Methodology

### *Research structure*

Bibliometric analysis has been used in the nursing field to analyze meta-approaches and data research (Kokol, 2021). Nonetheless, only one bibliometric analysis, including some ideas about the sustainability in the health sector and the role of nurses, is available (White et al., 2014). No recent study presents the current research lines reflecting new paradigms or interventions to achieve environmentally sustainable health systems. Therefore, the current research was structured to cover this topic following the workflow described by Aria and Cuccurulo (Aria & Cuccurullo, 2017): first, the research design, which included the research questions (RQs according to (Zupic & Čater, 2015); second, the selection of the bibliometric and visualization section (SPSS program and Vosviewer), next, the compilation of the bibliometric data (via three major databases in health: Scopus, Web of Sciences and PubMed), analysis (bibliometric analysis and use of programs), visualization and interpretation.

*Data gathering*

The research questions proposed were:

RQ1: Which are the publication trend and differences from the definitions of WHO?

RQ2: Which countries and journals contribute to this field, and what is their relationship?

RQ3: Which are the top publications and authors focused on interventions to obtain environmentally sustainable health systems and the inclusion of nurses?

RQ4: How has the research focus and major topics evolved in the timeframe?

RQ5: What influence of nurses have as agents and based on their workforce in the sustainability of health care systems?

Based on these RQs, the research strategy followed the population, intervention, comparison, and outcome structure (PICO), which led to selecting the keywords and Medical Subject Heading (MeSH). The research strategy was formed by search strings based on MeSH terms or keywords: “Sustainability", "Nursing" and "Environment" and a time limit of 10 years.

The review's inclusion criteria were articles indexed in the databases that contained some of the keywords from the thematic area of nursing. The exclusion criteria were other thematic areas and documents focused on other systems and healthcare workers and not framed in the WHO’s definition provided.

An initial search carried out in September 2021 using “nurse” and “sustainability” identified 1112 from Scopus and Web of Science (WOS) databases. The results were reduced after the revision from the researchers (O.M.L-A. and P.A-M), leaving 87 articles that would provide information on the topic. Based on the scarcity, the final research implemented in January 2022 was: TITLE-ABS-KEY (nurs* AND environment* AND sustainable) in Scopus; TS=( nurs* AND environment* AND sustainable) in Web of Sciences, and ((nurs*[Other Term]) AND (sustainable[Other Term])) AND (environment[MeSH Terms]) in PubMed.

After applying the time limit, 852 documents were obtained in Scopus, 677 papers in WOS, and 19 in PubMed, exported in an excel and bibliographic format (csv. and enw.) to be reviewed in the Endnote program (Clarivate Analytics, London, UK). The documents’ details included author(s), affiliation, type of publication, title, abstract, keywords, year of publication, language, and the number of citations. Additionally, to these results from the three databases (Scopus, WOS, and PubMed), further research was implemented in another database (Google Scholar and Dialnet) using the exact search string, and grew literature relevant to the topic (conference papers) was included.

The selection of the documents for the quantitative analysis followed the PRISMA recommendations (Page et al., 2021). Two researchers screened the documents' titles, abstracts, and keywords. During the analysis, 468 documents were eliminated since they focused on aspects unrelated topics, such as fishing, biodegradation of waste, or microplastic in the oceans. Also, 450 documents were not including since focused on the environmentally sustainable systems. Finally, 159 documents related to the nurses and sustainability in the health system environment (Figure 1).

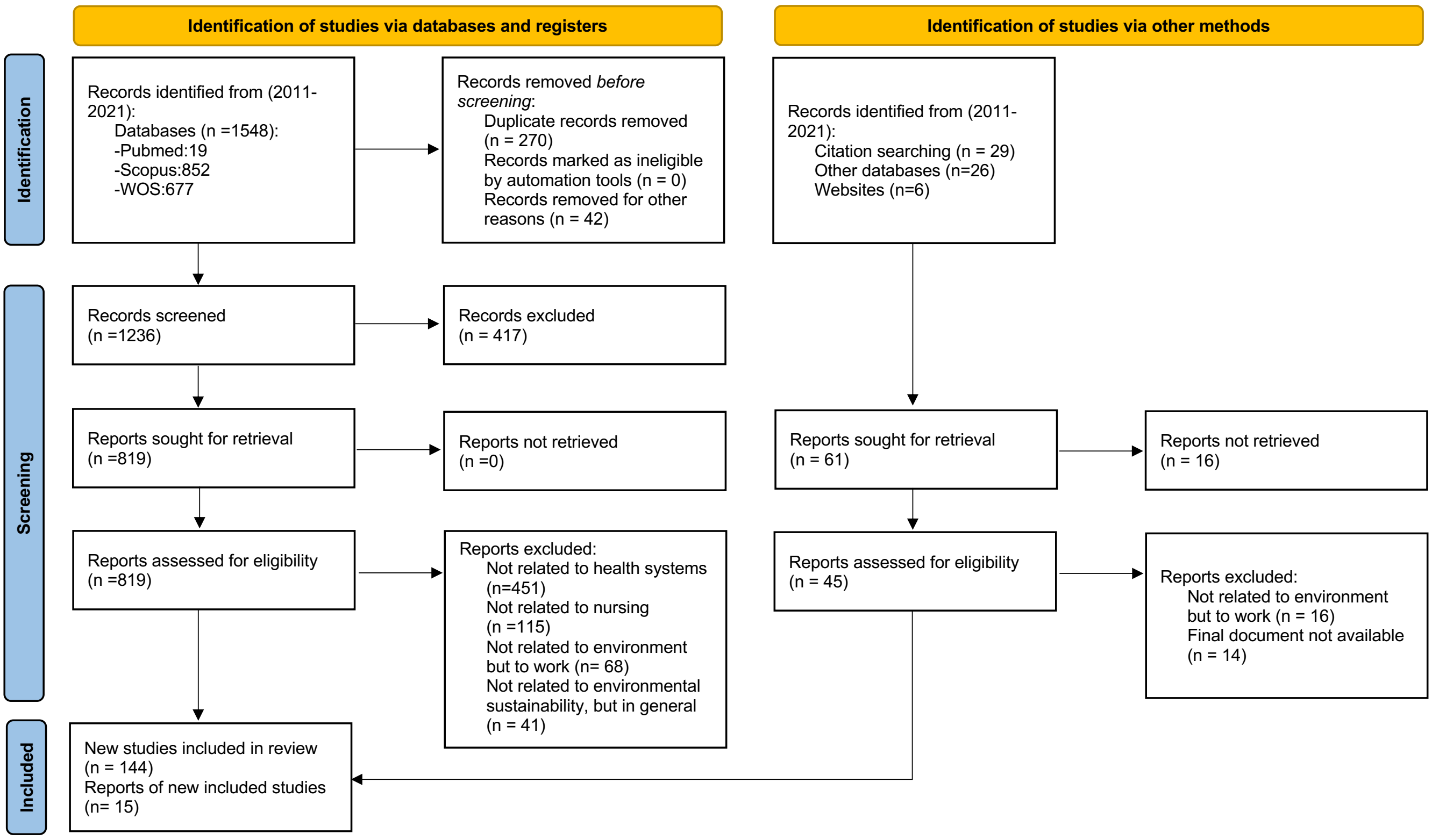

Figure 1. PRISMA 2020 flow diagram for systematic reviews (Page et al., 2021)

*Trend and Association Analysis*

After obtaining all the data structured in the csv. format using the Excel version 17 (Microsoft Corporation, Redmond, Washington, USA), SPSS program version 28 (IBM Corporation, Armonk, NY, USA), and VOSviewer version 1.6.15 (Ness Jan van Eck, Netherlands) were used to determine significant differences, citation analysis, and mapping and networking.

The data analysis was qualitative (mapping the items and checklist of the publications) and quantitative (statistical analysis). The qualitative analysis of the maps to identify the thematic and semantic structure of the scientific domain, visualizing its relationships with other keywords, completed with a manual and critical selection for the final filtering of the keywords, eliminating those that had to be with different themes such as stressful work environments or burnout. The checklist implemented was the Preferred Reporting Items for Systematic Reviews and Meta-Analyses (PRISMA), Strengthening the Reporting of Observational Studies in Epidemiology (STROBE), and Enhancing the Quality and Transparency Of health Research (EQUATOR).

The quantitative analysis included metrics (Journal Citation Report, quartile, and Journal Citation Indicator of the year 2020 and the year of publication) and details of the documents, such as the count of cites according to PlumX Metrics. The research results were initially analyzed using descriptive analysis, such as the frequencies of publications per country. The non-parametric tests (Kolmogorov-Smirnov test $p < 0.001$) were used accordingly to the variable, such as the Mann-Whitney test or Spearman's correlation.

**Results**

Table 1 showed the frequency of academic publications related to the environmentally sustainable health sector, including the role of nurses. The trend of publications indicated that the change occurred in 2017, being the most prolific researchers in 2019 and 2020. Additionally, the median of the year of publication was set in 2018, which matched the more substantial number of publications. From 2011 to 2021, the most published documents were articles (originals 81.10% and reviews 5.69%) being no significant differences according to the year of publication ($p$>0.05). These documents were mainly from the United States (32.08%), followed by the United Kingdom (10.69%). They were most of them published in indexed journals (Table 1), being independent of the year of publication ($p$>0.05). This trend of publication seemed to be associated with the number of citations ($p$<0.001), being more relevant than the difference between the 2017-2021 (value=44.49; $p$=0.028), and the JCR of the year of publication ($p$=0.046).

**Table 1.** The trend of publication on this topic and differences regarding the type of publication, the affiliation of the author, indexed at JCR and quartile (RQ1)

| YEAR OF PUBLICATION | FREQUENCY | TYPE OF DOCUMENTS | COUNTRY | INDEXED AT JCR | QUARTILE |
|---|---|---|---|---|---|
| **2011** | 1.3% | | | | |
| **2013** | 1.3% | | | | |
| **2014** | 3.8% | | | | |
| **2015** | 5.0% | | | | |
| **2016** | 8.8% | Articles 86.8%<br>*p*=0.91 | The United States of America (USA)<br>32.1%<br>*p*=0.28 | Indexed<br>61.0%<br>*p*=0.17 | Quartile (Q1-Q4)<br>61.6%<br>*p*=0.28 |
| **2017** | 13.1% | | | | |
| **2018** | 8.8% | | | | |
| **2019** | 15.6% | | | | |
| **2020** | 24.4% | | | | |
| **2021** | 18.1% | | | | |

The countries with a higher number of publications and higher number of citations were the UK (number of publications=10.7% and 8.25 ±14.29 cites; IC at 95% 0.64-15.68), USA (number of publications=32.1%; 5.92 ±6.98; IC at 95% 3.95-7.89), Australia (number of publications=8.8%; 3.78 ±3.31; IC at 95% 1.86-5.69), Spain (number of publications=6.9%; 6.0 ±9.49; IC at 95% 0.38-12.38) and Sweden (number of publications=5.0%; 11.25 ±17.87; IC at 95% -3.69-26.19). The co-currency of countries indicated that there were four clusters formed by nine countries (Figure 2), being the 1st (red) constituted by the USA (48 documents and eight links) and Sweden (9 documents and eight links). The second (green) was constituted by Australia (20 documents and eight links) and Taiwan (3 documents and three links). Meanwhile, the third was formed by the UK (23 documents and seven links) and Spain (16 documents and seven links), and the fourth was formed only by China. The comparison between the countries and associations between these was related to the number of cites ($p$=0.006), being more significant the differences between the countries with fewer publications, such as France (2.5%), compared to the USA (value=36.34; $p$=0.001), which was also associated with the JCI in 2020 ($p$=0.021).

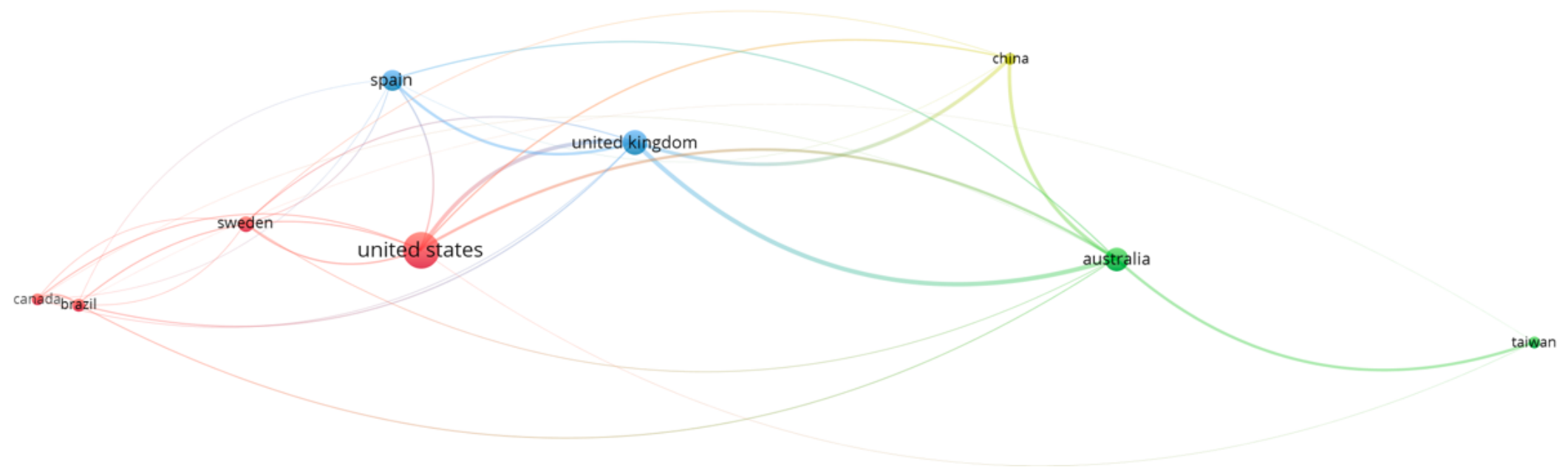

**Figure 2.** Co-currency of countries based on bibliographic coupling with a maximum of 25 documents and a minimum of 5 (RQ2).

According to the cites, the relevance of these countries in this topic is also reflected by the top ten articles (Table 2). Table 2 shows how the most relevant investigations (7/articles were published in the countries (Figure 2), which were also co-writing the results (3/10). Most of the studies were reviews (6/10), from systematics and scoping to scientometrics, followed by original studies, qualitative (2/10), and observational (2/10). The quality of the studies based on the checklists indicated that the qualitative analysis had higher methodological quality than compared to the observational studies (Anåker et al., 2015 had 93.75% meanwhile Richardson et al., 2014 had 34.38%). The data (Table 2) indicated that the most relevant articles were published in 2013 or 2014, being their thematic area focused on education, SDGs, and care. Only one of the top ten articles was in sync with the definition of an environmentally sustainable care system (Dossey et al., 2019), although other authors indicated the relevance of sustainability in health systems, including hospitals or primary care.

**Table 2.** The top ten of the most cited documents

| Title | Country | Type of study | Sample | Variables | Results | Source | Citations | Checklist |
|---|---|---|---|---|---|---|---|---|
| **Sustainability in nursing: A concept analysis** (Anåker & Elf, 2014) | Sweden | Scoping Review | 14 articles from 1990-2012 | Sustainable concept and healthcare sector related to the role of nursing | The research, through concept analysis, identified six attributes of sustainability and nursing. The results highly that the topics based on nursing education regarding ecology. The key to achieving it is through nursing academic programs. Additionally, the | *Scandinavian Journal of Caring Sciences (Indexed in JCR)* | 51 | Not applicable |
| | | | | | article emphasizes the relevance of including sustainability in healthcare organizations. | | | |
| **Primary health care and the Sustainable Goals Development** (Pettigrew et al., 2015) | Intercontinental (UK, Brazil, Belgium Ghana and Australia) | Comment | None | None | Primary health care has. A key role in achieving SDGs being indispensable strategies to adapt the working environment Also, it is highlighted the relevance of the workforce including | *The Lancet (Indexed in JCR)* | 51 | Not Applicable |
| | | | | | nurses or midwives, but there is a scarcity of a proposed strategy for implementation and its monitoring. | | | |
| **The use of evidence-informed sustainability scenarios in the nursing curriculum: Development and evaluation of teaching methods** | UK | Observational study | Thirty nursing students studying from the second year. | Skill facilitator and environmental awareness related to waste production | The results indicated education improves students' perception regarding the problem of the sustainability. The way to achieve awareness is through clinically relevant | *Nurse Education Today (Indexed in JCR)* | 35 | STROBE *11/32 (34.38%)* |
| (Richardson et al., 2014) | | | | | scenarios in skill sessions. | | | |
| **Nurses' perceptions of climate and environmental issues: A qualitative study** (Anåker et al., 2015) | Sweden | Qualitative study | Nurses (N = 18) were recruited from hospitals, primary care and emergency medical services. Eight participated in semi-structured, in-depth | Nurses' perceptions regarding climate and environmental issues and examine how nurses perceive their role in the creation of sustainable health care | Four areas were identified: incongruence between climate and environmental issues and nurse's daily work; and public health work is regarded as a health co-benefit of climate change mitigation. | *Journal of Advanced Nursing (Indexed in JCR)* | 33 | *EQUATOR 30/32 (93.75%)* |
| | | | individual interviews and 10 participated in two focus groups | | | | | |
| **Planetary Health and the Role of Nursing: A Call to Action** (Kurth, 2017) | USA | Literature Review | Not identified | Planetary health framework | The impact of health sector and workforce in the energy and water consumption, highlighting the need of sustainable programs in health care environments, especially renewable energies | *Journal of Nursing Scholarship (Indexed in JCR)* | 32 | Not Applicable |
| **Tweet if you want to be sustainable: A thematic analysis of a Twitter chat to discuss sustainability in nurse education** (Richardson, Grose, et al., 2016) | UK and Spain | Qualitative study: constructive paradigm | One hundred and nineteen people posted nine hundred nine hundred and ninety-six Tweets, with reach up to 3,306,368, was analysed | Social media as factor to engage nurses' awareness and perception | The analysis of the tweets highlighted the relevance of sustainability for nurses. The most important topics identified were the sustainability among nursing/nurses via education, with focused on waste, especially plastic | *Journal of Advanced Nursing (Indexed in JCR)* | 28 | *EQUATOR* 27/32 (84.38%) |
| **Including sustainability issues in nurse education: A** | European (UK, Germany, Spain and Switzerland) | Observational | First year Students Four European | Sustainability Attitudes in Nursing Survey (SANS_2) | This survey indicated a high score, so it could be highly useful to use it the | *Nurse Education Today* | 26 | STROBE *20/32 (62.5%)* |
| **comparative study of first year student nurses' attitudes in four European countries** (Richardson, Heidenreich, et al., 2016) | | | Universities (N=916) UK n=450 Germany n=196 Spain n=124 Switzerland n=146 | questionnaire and Demographic characteristics | nurses' awareness. There were significant differences in sustainability awareness of students between European countries, showing that German nurses' students had higher scores. | *(Indexed in JCR)* | | |
| **Towards environmentally responsible nursing: A critical** | Finland | Systematic review | Selected papers (N = 11) | Environmental issues in nursing science | The papers identified nurses as environmentally responsible for sustainability in | *Journal of Advanced Nursing (Indexed in JCR)* | 23 | PRISMA 33/53 62.26% |

| | | | | | | | |
|---|---|---|---|---|---|---|---|
| **interpretive synthesis** (Kangasniemi et al., 2013) | | | | | diverse sectors. They need to be targeted as agents in environmental management and tools for practical environmental responsibility should be included and presents in the curriculum. | | |
| **How the nursing profession can contribute to** | USA | Scientometrics analysis | N= 296 documents identified in the topic | Millennium development goals (MDGs) and the role of | The results indicated how nurses' contribution, as professionals, has | *Nursing Management (Indexed in JCR)* 22 | Not Applicable |

| | | | | | | | | |
|---|---|---|---|---|---|---|---|---|
| **sustainable development goals** (Benton & Shaffer, 2016) | | | | nurses in policies | limited their contribution to the MDGs link to health Despite not being so active in MDGs, nursing has been more proactive in addressing the SDGs. | | | |
| **Nursing and the sustainable development goals: From Nightingale to now** (Dossey et al., 2019) | USA | Literature Review | Not identified | Nurses as agents to contextual the SDGs related to the care provided and the system | The results highlight the evolution from the identification of Florence Nightingale of the relevance of environment to the relevant of sustainability for the | *American Journal of Nursing (Indexed in JCR)* | 22 | Not Applicable |

| | | |
|---|---|---|
| | care of patients, including nurses and adapting the environment. | |

The top five articles regarding the cites (Table 3) were carried out in Sweden, International collaborations (UK, Brazil, Belgium, Ghana, and Australia), the UK, and the USA. These results were published from 2013 to 2019, being among most of the top ten from the year 2014. Moreover, this table also indicated how the top articles were from relevant journals, indexed in JCR, and the first and the second quartile linked to the nursing field, which was linked to the number of cites ($p<0.001$).

A further analysis was carried out based on the dominant authors on this topic (RQ4). The top five authors (Table 3) who focused on this subject during the last decade were also from the top countries and collaborated among them (Figure 3). The mean of the h-index of the top ten authors was 13.6±12.1, the mean of citation of 1040.7, and a mean of 65.8 documents. The top authors were from the USA (Rosa W. E. and Dossey B.M.), the UK (Grose J. and Richardson J.), and Sweden (Elf M. and Anåker A.), followed by Brazil (Cunha I.C.K.O. and Furukawa P.d.O.) and Canada (Beck D.M. and Marck P.B.).

Rosa W. E. tops this field with 11 documents during the last decade, with 123 publications mainly in the area of *Dental Practice; Delivery Of Health Care; Environmental Sustainability*, followed by Grose J., who also published in the same area as Rosa W. E. Nonetheless, the author with the h-index is Richardson J. (h-index of 46), followed by Elf M. with an h-index of 16, Marck P.B with an h-index of 16, and is the fourth, Rosa W. E. with an h-index of 12 (Table 3). The top that published on this topic started to publish in 2012 and 2014, being also connected among the authors (Figure 3).

**Table 3.** Top 10 authors published in the topic, with h-index, citations, and total publications

| Author | Publications on the topic | H-index | Total Citations | Total Publications | First Publication | Affiliation | Author ID |
|---|---|---|---|---|---|---|---|
| Rosa W. E | 11 | 12 | 511 | 123 | 2014 | United States | 56194379200 |
| Grose J. | 4 | 11 | 394 | 30 | 2012 | United Kingdom | 55226482400 |
| Richardson J. | 3 | 46 | 7094 | 200 | 1993 | United Kingdom | 35500478000 |
| Beck D.M. | 3 | 5 | 80 | 19 | 1998 | Canada | 15761985500 |
| Cunha I.C.K.O. | 3 | 9 | 323 | 79 | 1995 | Brazil | 7003935865 |
| Elf M. | 3 | 16 | 699 | 59 | 2001 | Sweden | 23008276300 |
| Dossey B.M. | 3 | 11 | 358 | 68 | 1993 | United States | 7004496398 |
| Anåker A. | 2 | 7 | 173 | 12 | 2004 | Sweden | 56056321800 |
| Furukawa P.d.O. | 2 | 3 | 37 | 6 | 2010 | Brazil | 36463381900 |
| Marck P.B. | 2 | 16 | 738 | 62 | 1993 | Canada | 6701740195 |

The connection of the authors, based on co-citations, indicated that the top authors were cited between them (Figure 3). The co-citation of the authors pointed out that there were clusters among the authors, being the 1st (red) constituted by 14 authors, led by Richardson J. (citations=38, links=25), Grose J. (citations=26, links=25), Anåker A., (citations=24, links=24) and Elf M. (citations=23, links=24). The 2nd cluster (green) is formed by 11 authors, led by Rosa W. E (citations=34, links=21) and Dossey B.M. (citations=34, links=19). The last cluster is formed by three authors, led by Haines, A. (citations=19, links=26), whose h-index is 85, but mainly published on *Climate Change; Rockefeller Foundation; Malnutrition,* including Gonzalez-Garcia, S. from Spain (12 citations and h-index 41). These results indicated that most connections were established between researchers from USA, UK, Sweden, and Spain (RQ4).

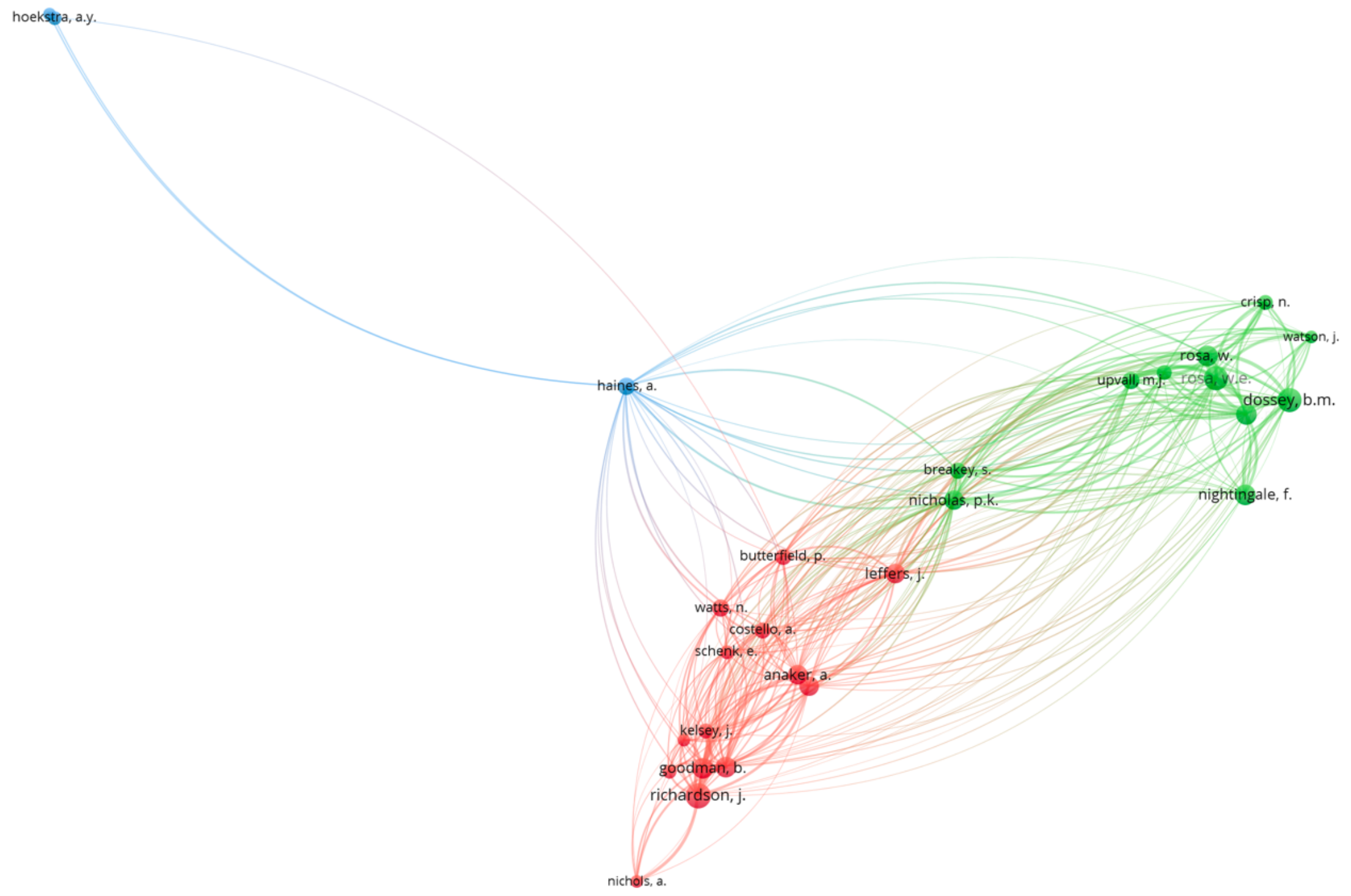

**Figure 3.** Connections between authors based on co-citation with 25 documents and a minimum of 10 (RQ4).

Finally, the keyword analysis was carried out to determine the current lines of research on this topic (RQ5). Four clusters were identified that include. Eighty keywords are represented in red (1$^{st}$ with 28 keywords, 1108 links, and 271 occurrences), green (2$^{nd}$ with 24 keywords, 875 links, and 182 occurrences), blue (3$^{rd}$ with 17 keywords, 687 links, and 261 occurrences), and the last in yellow (11 keywords, 476 links, and 110 occurrences) (Figure 4). The first cluster was led by the keywords "organization and management" (65 links and 20 occurrences) and "nursing education" (59 links and 23 occurrences). This cluster represented one of the main sub-topic topics based on sustainable education, including the training, curricula, and competencies. The second, including the terms “nurses” (71 links and 26 occurrences) and “waste management” (45 links and nine occurrences), highlighted the sub-topic of the nursing profession and nursing discipline in environmental sustainability. The cluster focused on primary health care responsible for preventing this impact through health promotion and prevention of diseases related to the environment environments and pollutants. The third cluster had as significant terms “sustainable development” (73 links and 52 occurrences) and “United Nations” (40 links and 25 occurrences). This cluster is based on environmental sustainability through protection and prevention of its impact, for example, through the waste that is generated or the essential elements of nature in the environmental sustainability of Florence Nightingale. The last cluster, whose leading keywords were “environmental protection” (61 links and 17 occurrences), “education” (51 links and 12 occurrences) and “organization” (46 links and ten occurrences), focused on health policy to maintain the environment. This cluster highlighted the relevance of education and policies to preserving the ecosystem and achieving environmental health.

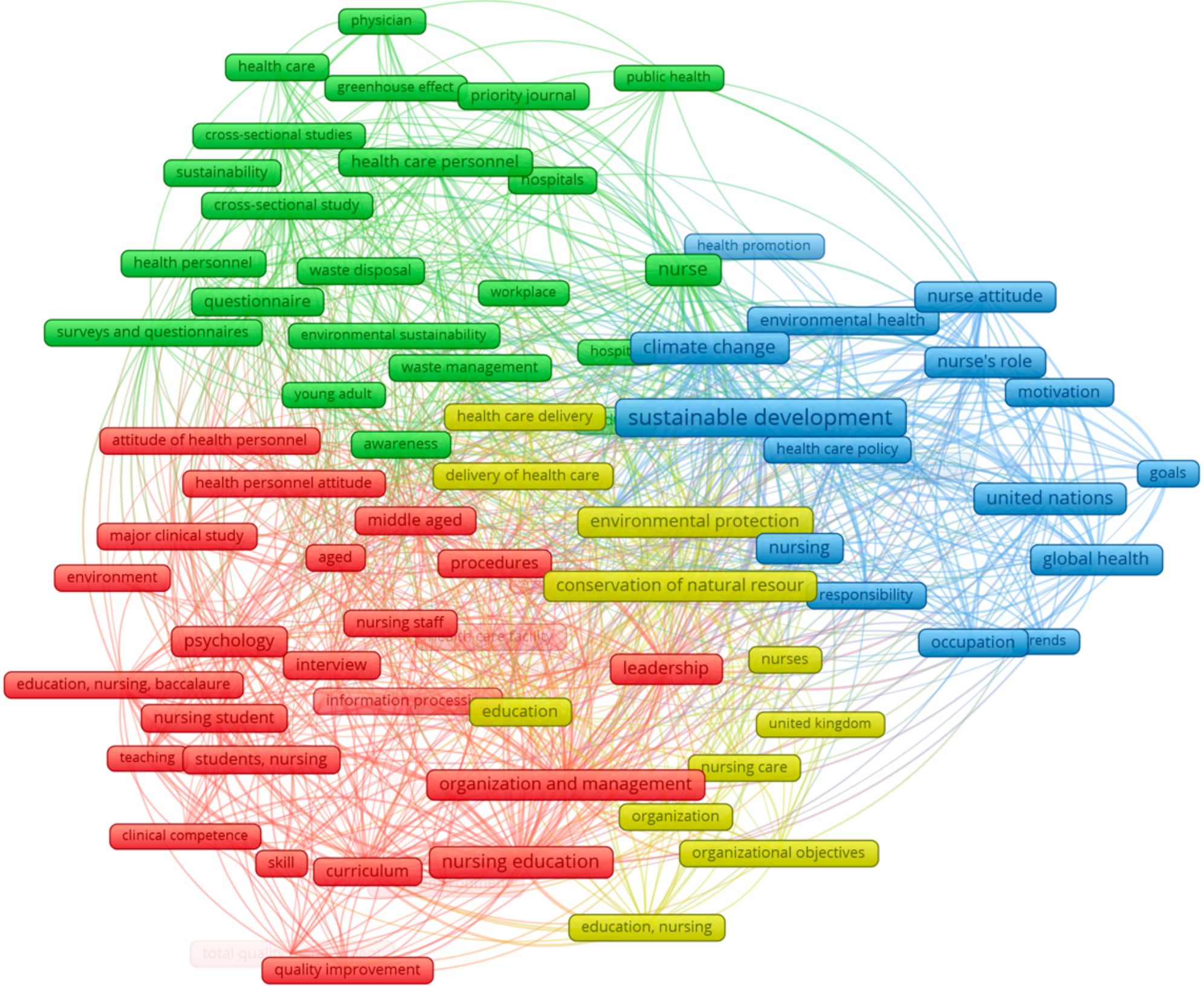

**Figure 4.** Co-occurrence of most common index terms per document with a minimum of 5 connections. Note: words unrelated to the topic were eliminated (human/s, article, gender, adult, and human experiments) (RQ5).

## Discussion

This study aimed to determine the current scientific knowledge and research lines focused on environmentally sustainable health systems, including the role of nurses during this last decade, and to choose the nursing education and interventions to improve the environmental awareness among nurses.

The trend of publication, major countries and connections between them, the most relevant authors, and what are the topics more analyzed by the authors indicated associations between countries, the importance of WHO's and United Nations' recommendations(General Assembly, 2011), and the most relevant topics.

First, the trend analysis indicated that the period with a higher number of publications is from 2017 to 2021, despite a slight decrease during 2021. This trend matched the inclusion of nursing in the sustainability (Benton & Shaffer, 2016) and WHO's definition of environmentally sustainable health systems (World Health Organization, 2017). This tendency also matches previous bibliometric analyses that indicated how nursing had grown exponentially during the last decade (Kokol, 2021). There are more reviews on this topic (Lilienfeld et al., 2018). Nonetheless, the Covid-19 has impacted this research since there was a decrease from 2020 to 2021, which could be explained by the findings of Osingada and Porta (2020). Their results stated how the pandemic had impacted this field and even reduced the number of available publications.

Besides, most of the publications were from Northern countries, interconnected among themselves. Based on several publications and citations, the most relevant countries were the USA, the UK, and Sweden; the top ten articles and authors collaborated among them. These results are consistent since major these countries are the geographical base of diverse international organizations, such as the United Nations.

The analysis of the top ten publications, mainly carried out in such countries and by the ten top authors like Anåker or Richardson, focused on the role of nursing in diverse areas but was highly repetitive about the relevance of the education. Also, it was reflected the lack of original studies framed on the WHO's definition (World Health Organization, 2017) and the role of nurses in obtaining it, which is understandable since the definition is recent. Despite that, the articles introduce the idea of environmentally sustainable health systems and the role of nursing through education and improving awareness but miss empirical data on how to achieve it. In this sense, the only article published by a significant author in the area (Dossey et al., 2019) and that identified nurses are relevant but does not present a unified action to achieve, only that education is critical.

In this sense, Yakusheva et al. (2022) highlight the need for value-informed decision-making to achieve environmentally sustainable health care systems. Only one of the articles, even published after 2017 (Dossey et al., 2019), presented the idea. This article also highlighted that achieving is through education matches the findings of articles and reviews analyzed (Lilienfeld et al., 2018; Rosa et al., 2019). Most of the top ten articles indicated the high relevance of education nurses as agents to integrate and carry out the SDGs and create an environmentally sustainable health system. These findings were also present in identifying the topic of publications shown by the concurrency of keywords. Moreover, from the analysis of the keywords, it can be concluded that the role of nursing in this topic is diverse but that it focuses on education, organization, and management, policies to maintain or create an environmentally sustainable system. Also, the analysis indicated that the organizations, mainly United Nations, and their reports or recommendations, like the SDGs, are essential to nurses as guides to incorporate actions and measures to be effective agents, which was also patent in different works of the ten top publications (Benton & Shaffer, 2016; Kurth, 2017; Pettigrew et al., 2015).

*Limitations and implications to the field*

The study's main limitation is the selection of the keywords, which was tried to mitigate by the inclusion of three databases and a follow-up of the PRISMA declaration carried out via a double peer screening. Additionally, the analysis was focused on the quantitative analysis, reducing the qualitative results to the top ten articles.

Despite these results, the findings are exciting since this is the first bibliometrics analysis to identify the role of nursing in achieving environmentally sustainable health care systems. The findings have two significant implications for the field. First, there is a need for further original publications in this era, which will increase exponentially in the next decade. The primary way to achieve this is through education via institutions and policies.

## Conclusion

The bibliometric analysis indicated that research in environmentally sustainable health care systems is currently more theoretical. The research literature told how nurses are pivotal to the environment. Still, there is a lack of publications that analysis on this topic. Nursing education is key to achieving WHO's definition, being relevant in it, organizations and management and policies, being more relevant than even high quality of education so nurses can be active and positive agents in creating and maintaining research in environmentally sustainable health care systems.

Figure 1

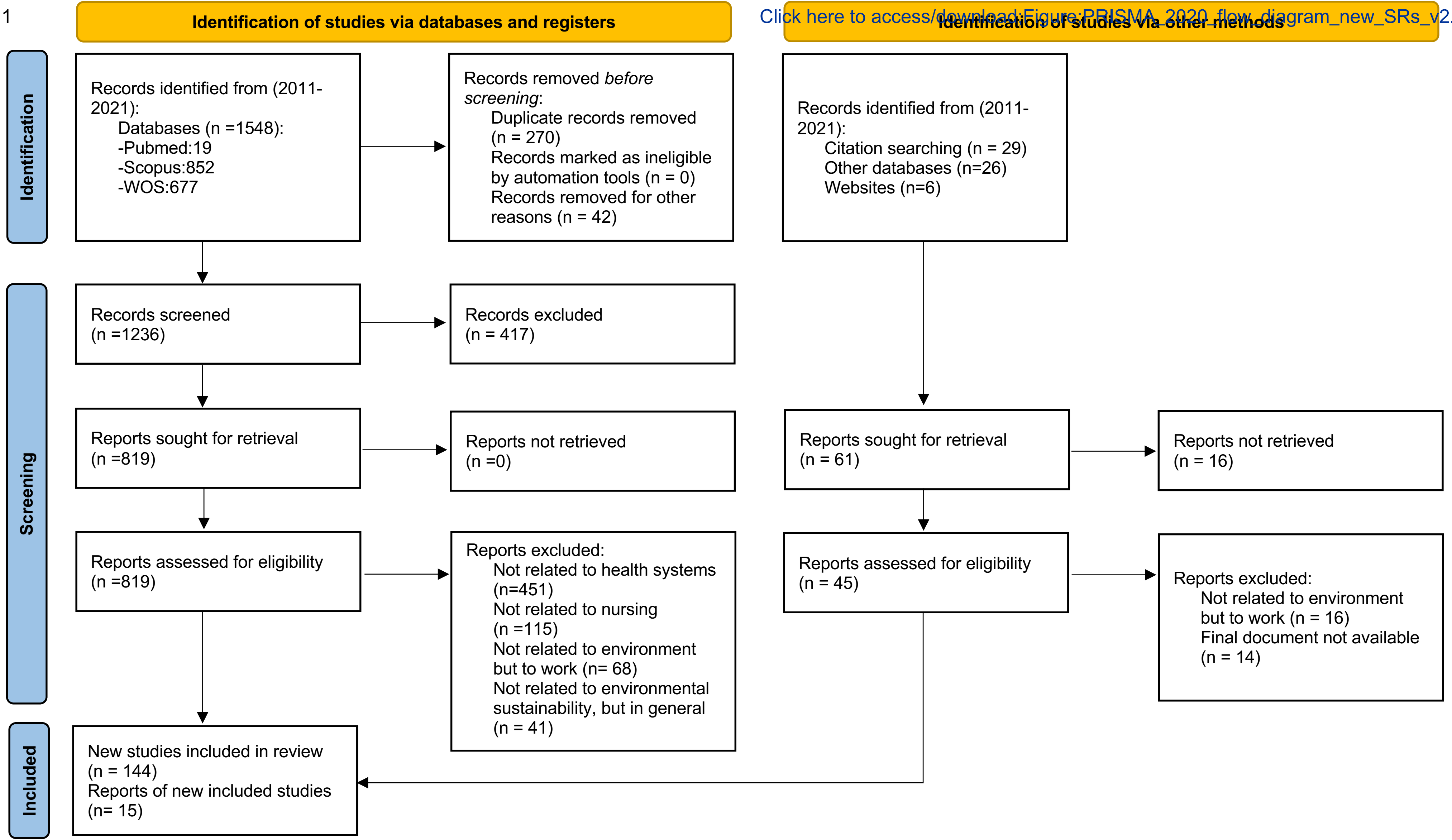
Identification of studies via databases and registers
Identification of studies via other methods
Identification
Screening
Included
Records identified from (2011-2021):
Databases (n =1548):
-Pubmed:19
-Scopus:852
-WOS:677
Records removed *before screening*:
Duplicate records removed (n = 270)
Records marked as ineligible by automation tools (n = 0)
Records removed for other reasons (n = 42)
Records identified from (2011-2021):
Citation searching (n = 29)
Other databases (n=26)
Websites (n=6)
Records screened (n =1236)
Records excluded (n = 417)
Reports sought for retrieval (n =819)
Reports not retrieved (n =0)
Reports sought for retrieval (n = 61)
Reports not retrieved (n = 16)
Reports assessed for eligibility (n =819)
Reports excluded:
Not related to health systems (n=451)
Not related to nursing (n =115)
Not related to environment but to work (n= 68)
Not related to environmental sustainability, but in general (n = 41)
Reports assessed for eligibility (n = 45)
Reports excluded:
Not related to environment but to work (n = 16)
Final document not available (n = 14)
New studies included in review (n = 144)
Reports of new included studies (n= 15)

Figure 2

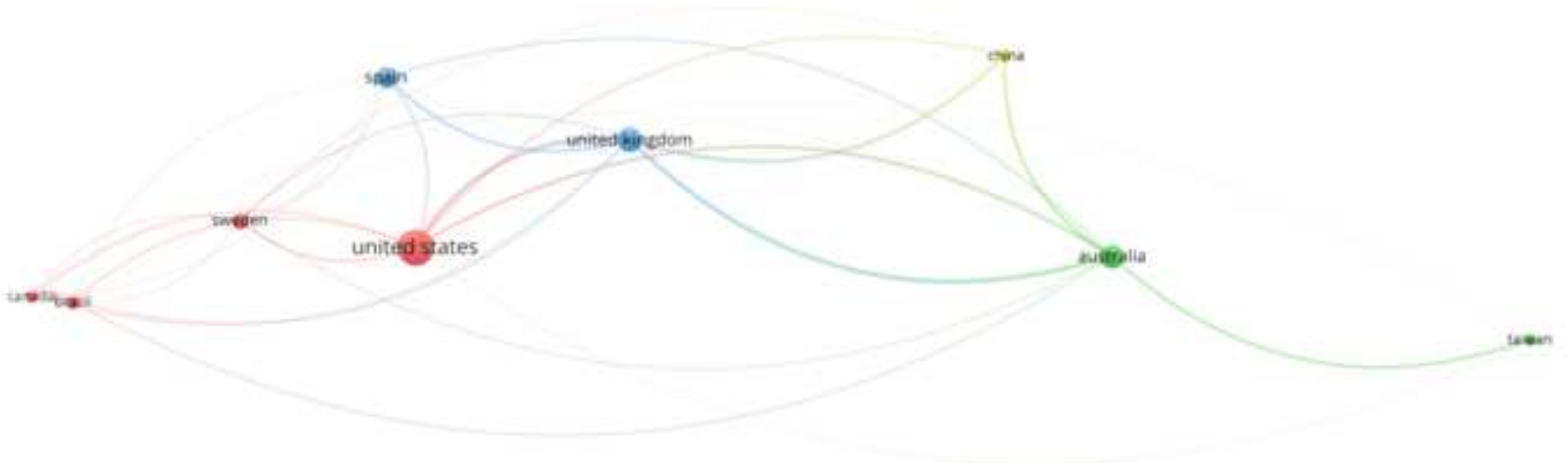

Figure 3

Click here to access/download;Figure;Figure 2.png

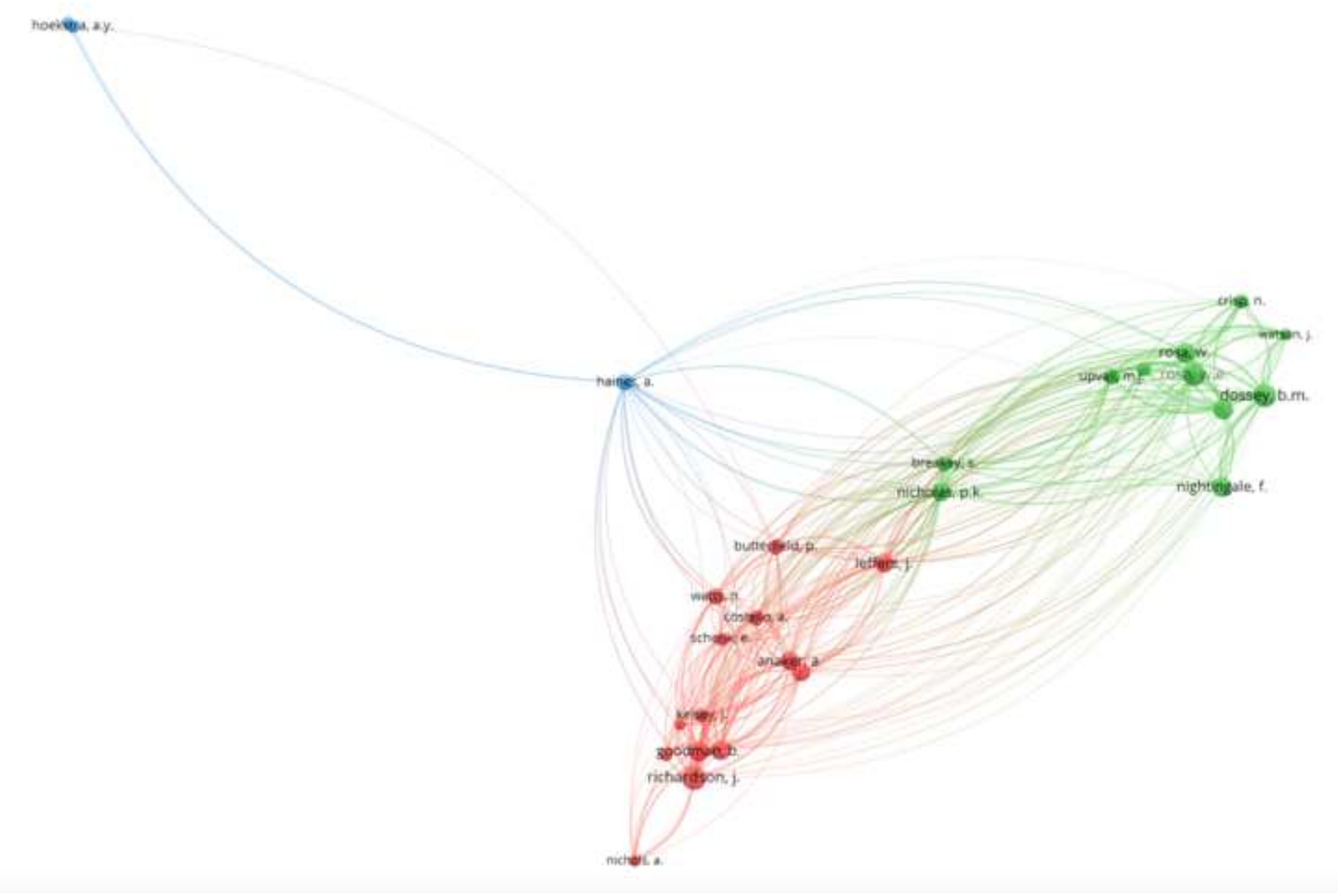

Figure 4

Click here to access/download;Figure;Figure 3.png

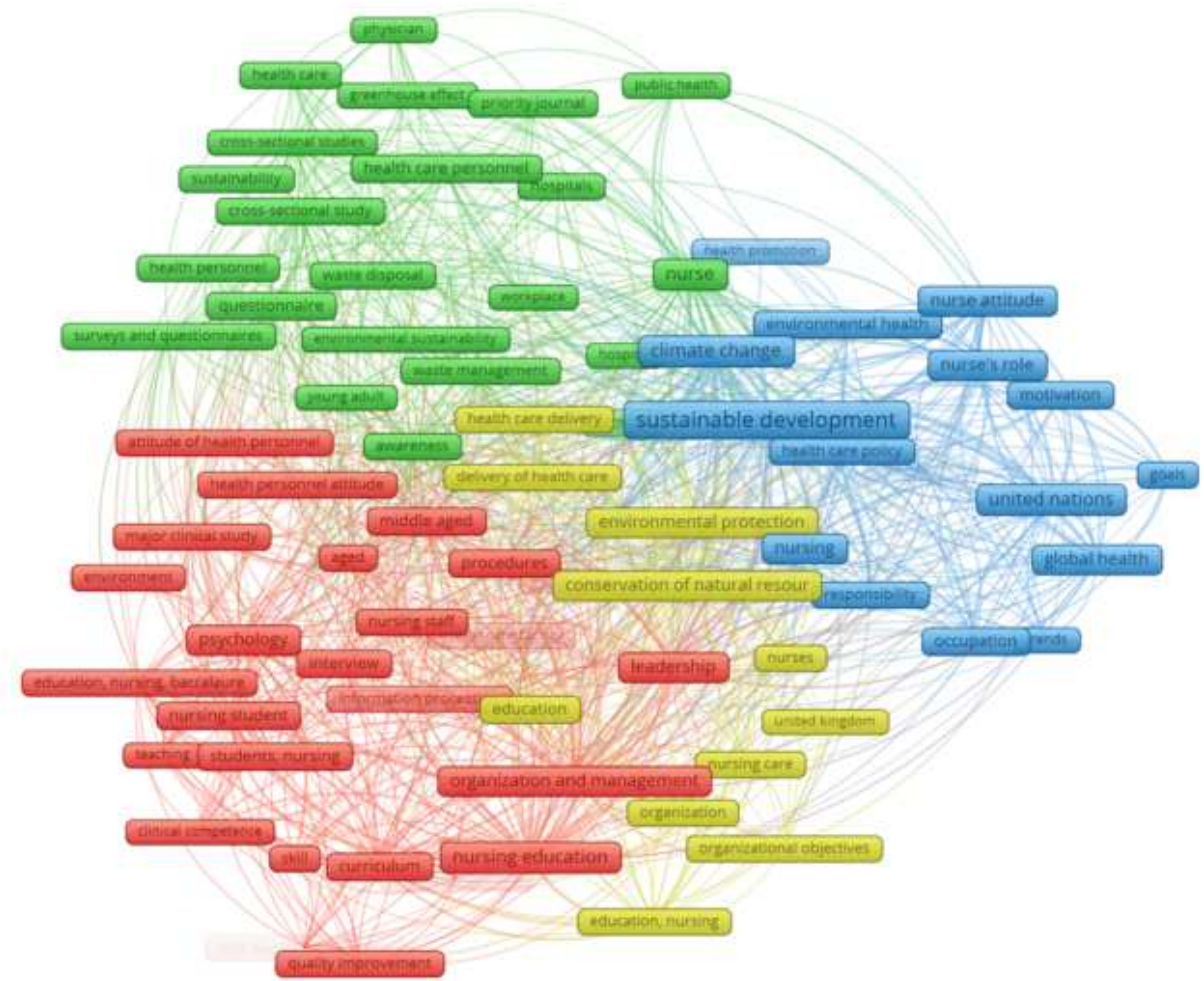

**Table 1.** The trend of publication on this topic and differences regarding the type of publication, the affiliation of the author, indexed at JCR and quartile (RQ1)

| YEAR OF PUBLICATION | FREQUENCY | TYPE OF DOCUMENTS | COUNTRY | INDEXED AT JCR | QUARTILE |
|---|---|---|---|---|---|
| **2011** | 1.3% | | | | |
| **2013** | 1.3% | | | | |
| **2014** | 3.8% | | | | |
| **2015** | 5.0% | | | | |
| **2016** | 8.8% | Articles 86.8% *p*=0.91 | The United States of America (USA) 32.1% *p*=0.28 | Indexed 61.0% *p*=0.17 | Quartile (Q1-Q4) 61.6% *p*=0.28 |
| **2017** | 13.1% | | | | |
| **2018** | 8.8% | | | | |
| **2019** | 15.6% | | | | |
| **2020** | 24.4% | | | | |
| **2021** | 18.1% | | | | |

**Table 2.** The top ten of the most cited documents

| *Title* | *Country* | *Type of study* | *Sample* | *Variables* | *Results* | *Source* | *Citations* | *Checklist* |
|---|---|---|---|---|---|---|---|---|
| ***Sustainability in nursing: A concept analysis*** (Anåker & Elf, 2014) | *Sweden* | Scoping Review | *14* articles from 1990-2012 | Sustainable concept and healthcare sector related to the role of nursing | *The research, through concept analysis, identified six attributes of sustainability and nursing. The results highly that the topics based on nursing education regarding ecology. The key to achieving it is through nursing academic programs. Additionally, the article emphasizes* | *Scandinavian Journal of Caring Sciences (Indexed in JCR)* | 51 | *Not applicable* |

| | | | | | | | |
|---|---|---|---|---|---|---|---|
| | | | | | *the relevance of including sustainability in healthcare organizations.* | | |
| ***Primary health care and the Sustainable Goals Development*** (Pettigrew et al., 2015) | *Intercontinental (UK, Brazil, Belgium Ghana and Australia)* | *Comment* | *None* | *None* | *Primary health care has. A key role in achieving SDGs being indispensable strategies to adapt the working environment Also, it is highlighted the relevance of the workforce including nurses or midwives, but there is a scarcity* | *The Lancet 51 (Indexed in JCR)* | Not Applicable |
| | | | | | *of a proposed strategy for implementation and its monitoring.* | | |
| ***The use of evidence-informed sustainability scenarios in the nursing curriculum: Development and evaluation of teaching methods*** (Richardson et al., 2014) | *UK* | *Observational study* | *Thirty nursing students studying from the second year.* | *Skill facilitator and environmental awareness related to waste production* | *The results indicated education improves students' perception regarding the problem of the sustainability. The way to achieve awareness is through clinically relevant scenarios in skill sessions.* | *Nurse Education Today (Indexed in JCR)* 35 | STROBE *11/32 (34.38%)* |

| | | | | | | | | |
|---|---|---|---|---|---|---|---|---|
| ***Nurses' perceptions of climate and environmental issues: A qualitative study*** (Anåker et al., 2015) | *Sweden* | *Qualitative study* | Nurses (N = 18) were recruited from hospitals, primary care and emergency medical services. Eight participated in semi-structured, in-depth individual interviews and 10 participated in two focus groups | Nurses' perceptions regarding climate and environmental issues and examine how nurses perceive their role in the creation of sustainable health care | Four areas were identified: incongruence between climate and environmental issues and nurse's daily work; and public health work is regarded as a health co-benefit of climate change mitigation. | *Journal of Advanced Nursing (Indexed in JCR)* | *33* | *EQUATOR 30/32 (93.75%)* |
| ***Planetary Health and the Role of Nursing: A Call to Action*** (Kurth, 2017) | *USA* | *Literature Review* | *Not identified* | Planetary health framework | The impact of health sector and workforce in the energy and water consumption, highlighting the need of sustainable programs in health care environments, especially renewable energies | *Journal of Nursing Scholarship (Indexed in JCR)* | *32* | Not Applicable |
| ***Tweet if you want to be sustainable: A thematic analysis of a Twitter chat to*** | UK *and Spain* | *Qualitative study: constructive paradigm* | *One hundred and nineteen people posted nine hundred and ninety-six* | *Social media as factor to engage nurses' awareness and perception* | The analysis of the tweets highlighted the relevance of sustainability for nurses. The most | *Journal of Advanced Nursing (Indexed in JCR)* | *28* | *EQUATOR* 27/32 (84.38%) |
| ***discuss sustainability in nurse education*** (Richardson, Grose, et al., 2016) | | | *Tweets, with reach up to 3,306,368, was analysed* | | important topics identified were the sustainability among nursing/nurses via education, with focused on waste, especially plastic | | | |
| ***Including sustainability issues in nurse education: A comparative study of first year student nurses' attitudes in four European countries*** | *European (UK, Germany, Spain and Switzerland)* | Observational | *First year Students Four European Universities (N=916) UK n=450 Germany n=196 Spain n=124* | *Sustainability Attitudes in Nursing Survey (SANS_2) questionnaire and Demographic characteristics* | *This survey indicated a high score, so it could be highly useful to use it the nurses' awareness. There were significant differences in sustainability awareness of students between European* | *Nurse Education Today (Indexed in JCR)* | *26* | STROBE *20/32 (62.5%)* |
| (Richardson, Heidenreich, et al., 2016) | | | *Switzerland n=146* | | *countries, showing that German nurses' students had higher scores.* | | | |
| ***Towards environmentally responsible nursing: A critical interpretive synthesis*** (Kangasniemi et al., 2013) | Finland | Systematic review | Selected papers (N = 11) | Environmental issues in nursing science | The papers identified nurses as environmentally responsible for sustainability in diverse sectors. They need to be targeted as agents in environmental management and tools for practical environmental responsibility should | *Journal of Advanced Nursing (Indexed in JCR)* | *23* | PRISMA 33/53 62.26% |

| | | | | | | | |
|---|---|---|---|---|---|---|---|
| | | | | | be included and presents in the curriculum. | | |
| ***How the nursing profession can contribute to sustainable development goals*** (Benton & Shaffer, 2016) | USA | Scientometrics analysis | N= 296 documents identified in the topic | Millennium development goals (MDGs) and the role of nurses in policies | The results indicated how nurses' contribution, as professionals, has limited their contribution to the MDGs link to health Despite not being so active in MDGs, nursing has been more proactive in addressing the SDGs. | *Nursing Management (Indexed in JCR)* | *22* | Not Applicable |
| ***Nursing and the sustainable development goals: From Nightingale to now*** (Dossey et al., 2019) | USA | *Literature Review* | *Not identified* | Nurses as agents to contextual the SDGs related to the care provided and the system | The results highlight the evolution from the identification of Florence Nightingale of the relevance of environment to the relevant of sustainability for the care of patients, including nurses and adapting the environment. | *American Journal of Nursing (Indexed in JCR)* | *22* | Not Applicable |

**Table 3.** Top 10 authors published in the topic, with h-index, citations, and total publications

| Author | Publications on the topic | H-index | Total Citations | Total Publications | First Publication | Affiliation | Author ID |
|---|---|---|---|---|---|---|---|
| Rosa W. E | 11 | 12 | 511 | 123 | 2014 | United States | 56194379200 |
| Grose J. | 4 | 11 | 394 | 30 | 2012 | United Kingdom | 55226482400 |
| Richardson J. | 3 | 46 | 7094 | 200 | 1993 | United Kingdom | 35500478000 |
| Beck D.M. | 3 | 5 | 80 | 19 | 1998 | Canada | 15761985500 |
| Cunha I.C.K.O. | 3 | 9 | 323 | 79 | 1995 | Brazil | 7003935865 |
| Elf M. | 3 | 16 | 699 | 59 | 2001 | Sweden | 23008276300 |
| Dossey B.M. | 3 | 11 | 358 | 68 | 1993 | United States | 7004496398 |
| Anåker A. | 2 | 7 | 173 | 12 | 2004 | Sweden | 56056321800 |
| Furukawa P.d.O. | 2 | 3 | 37 | 6 | 2010 | Brazil | 36463381900 |
| Marck P.B. | 2 | 16 | 738 | 62 | 1993 | Canada | 6701740195 |

**Author statement**: conception and design (A.G. and M. V-A), acquisition of data (O.M. L. and P.A-M.), or analysis and interpretation of data (O.M. L. and P.A-M.); drafting the article (O.M. L.); revising it critically for important intellectual content (P.A-M.); and final approval of the version to be published(O.M. L., P.A-M., A.G., and M. V-A,).